\RequirePackage{fix-cm}
\documentclass[twocolumn]{svjour3}

\usepackage{graphicx}
\usepackage{mathptmx}
\usepackage{siunitx}

\usepackage{balance}
\usepackage{float}
\usepackage{fancyhdr}
\usepackage{fnpos}
\usepackage{array}
\usepackage{setspace}
\usepackage[compact]{titlesec}
\usepackage{hyperref}
\usepackage{stfloats}
\usepackage{mathtools}

\usepackage{xcolor}

\journalname{Journal}

\begin{document}

\title{{\color{black}Dispersion control in pressure-driven flow through bowed rectangular microchannels}}
\author{Garam Lee\textit{$^{1}$} \and
Alan Luner\textit{$^{2}$} \and
Jeremy Marzuola\textit{$^{2}$} \and
Daniel M. Harris\textit{$^{1\ast}$}
}

\institute{
Daniel M. Harris\at
\email{daniel\_harris3@brown.edu}  \and
\textit{$^{1}$~School of Engineering, Brown University, Providence, RI 02912, USA.} \\ \\
\textit{$^{2}$~Department of Mathematics, University of North Carolina at Chapel Hill, Chapel Hill, NC 27599, USA.}
}
\date{Received: date / Accepted: date}

\maketitle

\begin{abstract}
In fully-developed pressure-driven flow, the spreading of a dissolved solute is enhanced in the flow direction due to transverse velocity variations in a phenomenon now commonly referred to as Taylor-Aris dispersion.  It is well understood that the characteristics of the dispersion are sensitive to the channel’s cross-sectional geometry.  Here we demonstrate a method for manipulation of dispersion in a single rectangular microchannel via controlled deformation of its upper wall. Using a rapidly prototyped multi-layer microchip, the channel wall is deformed by a controlled pressure source allowing us to characterize the dependence of the dispersion on the deflection of the channel wall and overall channel aspect ratio.  For a given channel aspect ratio, an optimal deformation to minimize dispersion is found, consistent with prior numerical and theoretical predictions.  Our experimental measurements are also compared directly to numerical predictions using an idealized geometry.  
\end{abstract}

\section{Introduction}
 When a localized patch of dissolved solute or other passive tracer is introduced into a pressure-driven flow, it spreads rapidly as it moves downstream through the subtle interplay of fluid advection and molecular diffusion \cite{4,aris1956}.  This phenomenon is now commonly referred to as Taylor-Aris dispersion and plays a critical role in many continuous microfluidic devices.  Depending on the application, it may be desirable to enhance dispersion (e.g. for mixing or reactions \cite{nagy2012mixing,handique2006}) or to minimize dispersion (e.g. for separations or chromatography \cite{dutta2003,westerbeek2020}).  In the present work, we develop a novel technique for experimental control of dispersion on a single microchip via the continuous deformation of a rectangular channel's side wall.  We first summarize the relevant physical quantities associated with the problem in what follows, also reviewed in detail within the 2006 review paper by Dutta et al. \cite{3dutta2006}.


Consider steady, fully-developed, pressure-driven flow through a channel of uniform cross-section.  At some instant in time, a localized patch of a passive solute is introduced into the flow and transported downstream and spreads due to the effects of advection and diffusion. On long time scales, the solute assumes a Gaussian shape and grows diffusely with an enhanced dispersion in the flow direction that can be described by a Taylor-Aris dispersivity $K$

\begin{equation}
    \frac{K}{\kappa} = 1 + \frac{1}{210}\left(\frac{Uh}{\kappa}\right)^2f = 1+\frac{1}{210} Pe^2f
    \label{eqn:dispersivity_arbitrary}
\end{equation}

\noindent where $\kappa$ is the molecular diffusivity of the solute, $U$ is the mean flow velocity, $h$ is the characteristic channel height, and $f$ is a scale-free parameter that depends only on the cross-sectional shape of the channel. The Peclet number $Pe=Uh/\kappa$ represents a ratio of the diffusive to advective time scales. The second term on the right hand side of equation (\ref{eqn:dispersivity_arbitrary}) represents the enhancement due to the presence of a non-uniform fluid flow and becomes dominant for large Peclet numbers.  

Over the past several decades, there has been persistent interest in understanding the role of a microchannel's cross-sectional shape on its dispersion characteristics \cite{16chatwin1981,7dutta2001,3dutta2006,ajdari2006,bontoux2006,14aubin2008,callewaert2014,yan2015chip,17aminian2016}. As defined in equation (\ref{eqn:dispersivity_arbitrary}), the cross-sectional shape defines the dispersion factor $f$ through the boundary conditions imposed on the flow problem.  Physically, the dispersion factor $f$ is directly correlated to how ``uniform'' the velocity profile is across a particular channel's cross-section.  A more uniform flow field will tend to lead to lower dispersion.

\begin{figure}
    \centering
    \includegraphics[width=84mm]{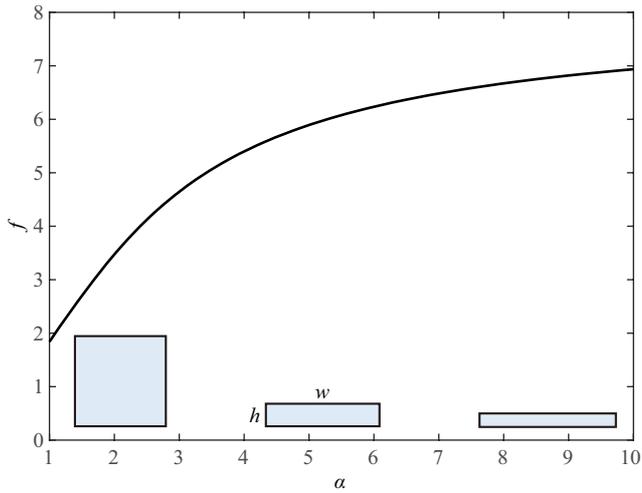}
    \caption{Dispersion factor $f$ of a rectangular channel as a function of the channel's aspect ratio $\alpha=w/h$.  The dispersion factor $f$ is defined in equation (\ref{eqn:dispersivity_arbitrary}) {\color{black} and computed numerically using the method described in Section 2 based on a standard theoretical formulation \cite{3dutta2006}}.  For reference, the case of infinite parallel plates corresponds to a dispersion factor of $f=1$.}
    \label{fig:1_Rect}
\end{figure}

The dispersion factor for a rectangular channel as a function of the aspect ratio $\alpha=w/h${\color{black}$\geq 1$} is shown in Figure \ref{fig:1_Rect}. As a point of reference, the case of two infinite parallel plates corresponds to a dispersion factor of $f=1$.  All rectangular geometries exceed this value due to the presence of the vertical side walls, {\color{black} even in the limit where $\alpha\rightarrow\infty$ (where $f\approx8$) \cite{doshi1978three}}.  More recently, numerical and theoretical work have focused on how practical modifications to the traditional rectangular channel geometry might reduce the dispersion factor \cite{7dutta2001,3dutta2006,callewaert2014}.  One particular message that arises throughout these works is that by preferentially enlarging the cross-section near the vertical side walls, it is possible to reduce the dispersion factor below the value for the pure rectangle.  This finding is rationalized physically in the review by Dutta et al. \cite{3dutta2006}: since a considerable amount of flow restriction (i.e. locally slower fluid flow) arises due to the presence of the vertical side walls, decreasing the local hydrodynamic resistance in these side regions leads to a more uniform velocity profile across the channel, thereby reducing overall dispersion.  Despite the numerical and theoretical progress in this direction, experimental results are lacking.

\begin{figure}
    \centering
    \includegraphics[width=84mm]{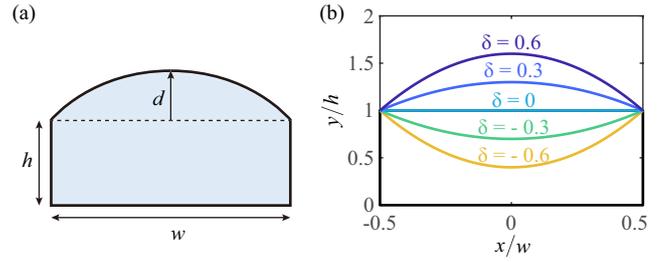}
    \caption{Cross-sectional geometry. (a) Geometrical parameters of the channel with deformation of one wall. (b) The deflection parameter $\delta$ is defined as the ratio of the maximum deformation $d$ over the original undeformed channel height $h$ as $\delta=d/h$. Positive $\delta$ represents outward deformation, and negative $\delta$ represents inward deformation.}
    \label{fig:2_DeltaDefine}
\end{figure}

The present work focuses on a single class of cross-sectional geometries shown in Figure \ref{fig:2_DeltaDefine}.  One wall of a rectangular channel of overall aspect ratio $\alpha$ is deformed in a continuous curve with a peak deformation $d$.  For this class of shapes, we expect the dispersion factor $f$ to depend predominantly on two dimensionless geometric quantities: the base aspect ratio $\alpha$ and deflection parameter $\delta=d/h$.  Note that a positive deflection parameter $\delta>0$ corresponds to an outward bowing of the channel wall, a negative deflection parameter $\delta<0$ refers to an inward bowing, and $\delta=0$ is the undeformed rectangle.  For all geometries studied, we define $h$ as the height of the undeformed rectangular geometry.

Dutta et al. \cite{3dutta2006} theoretically and numerically analyzed a similar ``bowed'' geometry (with a parabolic height profile on {\it both} top and bottom) and demonstrated that an optimal bowing exists to minimize the dispersion coefficient for a given channel aspect ratio.  In particular, in the asymptotic limit of $\left|\delta\right|\ll 1$ and $\alpha \gg 1$ it was predicted that this local minimum in the dispersion coefficient should occur at a deflection parameter of
\begin{equation}
    \delta_{o}=-1.654 \frac{h}{w} = -\frac{1.654}{\alpha}.
    \label{eqn:dopt}
\end{equation}
This asymptotic analysis can be applied to our idealized geometry as well, and yields the same result. According to their result, the optimal deflection parameter is predicted to be inwards ($\delta_o<0$) with its magnitude decreasing with the channel's aspect ratio.

To the best of the authors' knowledge, there remains no experimental validation of the prediction of an optimal bowing.  One recent experiment work relevant to the present discussion was contributed by Miles et al. \cite{miles2019} who demonstrated the possibility of ``flattening'' the velocity profile in a high-aspect ratio channel via bowing.  However only a single channel and deflection parameter were studied, and dispersion coefficients were not measured as part of this work.

The principal contributions of the present work are as follows.  First, we develop and describe a novel and accessible rapid-prototyping fabrication technique which allows for continuous control of a microchannel's cross-section by deforming one channel wall using a controlled external pressure source.  Dispersion experiments are then performed which demonstrate that the dispersion factor can be directly controlled on a single microchip by nearly an order of magnitude.  Finally, the measured dispersion factors are shown to be in good agreement with numerical predictions of an idealized geometry and both exhibit a local minimum in the dispersion factor at an optimal deflection parameter.





\section{Numerical method and prediction}

    The dispersion factor $f$ can be computed as sequence of two 2D Poisson problems over the cross-sectional domain of the channel. The theoretical formulation of this problem is reviewed in detail in prior work \cite{3dutta2006}, {\color{black} but we review the governing equations here briefly. We present the non-dimensional formulation of the problem which is normalized by characteristic length scale $h$ and characteristic velocity scale $p_z h^2/\mu$, where $p_z$ is the pressure gradient in the $z$ (flow) direction and $\mu$ is the fluid's dynamic viscosity. The first Poisson problem corresponds to the fully-developed steady flow problem through a channel of uniform cross-section $\Omega$:
\begin{equation}
    \nabla^2 u = -1
\end{equation}
with $u=0$ on the boundaries of the cross-sectional domain.  In this formula, $u(x,y)$ is the non-dimensional velocity field.  From this solution we can compute the average non-dimensional velocity $\bar{u}$ by integrating over the cross-section and normalizing by the cross-sectional area $|\Omega|$.  The second Poisson problem is
\begin{equation}
    \nabla^2 g = 1-\frac{u}{\bar{u}} \qquad \mathrm{and} \qquad \int_{\Omega}g\ dA  =0
\end{equation}
with $\nabla g \cdot \hat{n} =0$ on the boundaries of the domain, where $\hat{n}$ is the unit vector normal to the boundary.  We can then use the solution $g(x,y)$ to compute the dispersion coefficient $f$ as
\begin{equation}
    f(\Omega)=\frac{210}{|\Omega|}\int_{\Omega}\frac{u}{\bar{u}}g \ dA. 
\end{equation}

In order to solve these two problems numerically,} we parameterize the cross-sectional domain via a polygonal approximation made with a fine discretization of the boundary curves.  Then, we solve the corresponding Poisson problems through the MATLAB PDE solver package by generating a reasonably fine triangular mesh and the resulting finite-element stiffness matrices for both Dirichlet and Neumann boundary conditions.  From these, we are able to accurately integrate the numerical solution over the domain and compute the dispersion factor $f$ for an arbitrary cross-sectional geometry.

In the present work, the cross-sectional geometry is defined by two non-dimensional parameters: the aspect ratio $\alpha=w/h$ and the deflection parameter $\delta=d/h$.  For the purposes of the numerical calculation we assume the deformed boundary to take the form of a parabola, as plotted in Figure \ref{fig:2_DeltaDefine}(b).  {\color{black} For the idealized case of a thin membrane (i.e. negligible bending resistance) under a uniform pressure loading, the linearized governing equation predicts a parabolic profile \cite{ziebart1998mechanical}.  Despite this fact,} for small deformations the numerical results are found to be relatively insensitive to the exact functional form of the smooth upper curve.  

\begin{figure}
    \centering
    \includegraphics[width=84mm]{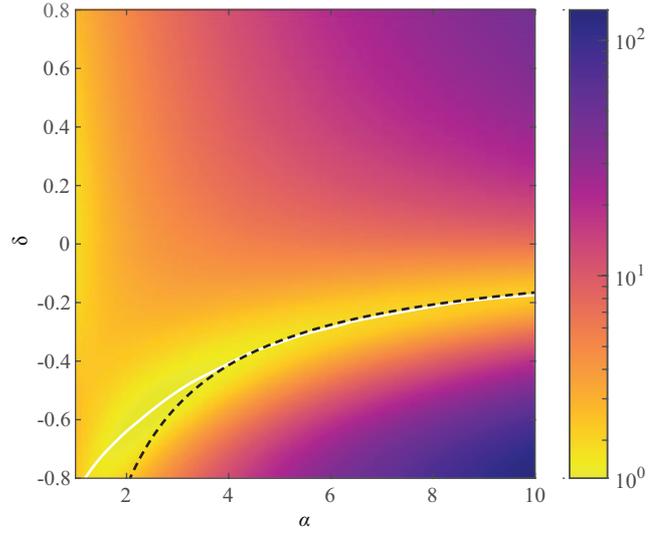}
    \caption{Numerical calculation of the dispersion factor $f$ as a function of the aspect ratio $\alpha=w/h$ and deflection parameter $\delta$. The white solid line indicates a local minimum in the numerically determined dispersion factor corresponding to an ``optimal'' value $\delta_o$ for each aspect ratio. The dashed line represents an asymptotic analytical prediction for such a minimum (Eq. (\ref{eqn:dopt}), \cite{3dutta2006}). }
    \label{fig:3_Numerical}
\end{figure}

The results of the numerical calculation are summarized in Figure \ref{fig:3_Numerical}.  Of particular note, is that for each aspect ratio an optimal deflection parameter $\delta_o$ exists at which the dispersion factor achieves a local minimum.  These results are consistent with the prior numerical work on bowed channels wherein both the top and bottom walls are deformed symmetrically \cite{3dutta2006}.  In the figure, we have also overlaid the asymptotic prediction for the optimal deflection parameter for a bowed channel (Eq. (\ref{eqn:dopt})) as predicted by Dutta \cite{3dutta2006} which shows good agreement with our numerical predictions, particularly for large aspect ratios, the regime in which this theoretical result is derived.

The numerical code used to perform these calculations is included as supplementary material.  The results of the numerical study will be compared directly to experimental measurements in a following section.  

\section{Experimental methods}

\subsection{Background}

In the field of microfluidics, a commonly used technique to fabricate a microchannel is soft lithography. Soft lithography uses a PDMS layer made by pouring an uncured PDMS mixture on a patterned mold.  This mold is generally a  produced by an optical lithography technique. The cured layer of PDMS is peeled off of the mold and bonded to a substrate using plasma cleaning. Despite the wide applications and advantages of this technique, the required expensive equipment, highly trained personnel, and cleanroom environment are not readily accessible to all researchers.  Furthermore the resources typically required for a single mold can limit early-stage prototyping of channel designs.


For our work, we use a fabrication method that relies on a commercial desktop craft cutter. In this technique, sometimes referred to as xurography, double-sided tape is cut into the designed channel shape by a commercial craft cutter and adhered to a transparent film and an acrylic sheet as an upper and lower boundary of the channel. Using a craft cutter for microchannel fabrication has been proposed before \cite{19yuen2009}; however, the lack of the necessary wall smoothness and resolution for many microfluidic experiments have limited the implementation of the technique. Recent work \cite{8taylor2019} demonstrated an optimized fabrication technique which achieved a surface roughness comparable to other leading fabrication techniques and can produce channels with dimensions as small as $\sim$ \SI{100}{\um}. Ease of use of the devices and extremely short fabrication time makes prototyping of different channel designs significantly less costly.  Furthermore, for controlled Taylor dispersion experiments, a channel with a much longer length than typical microchannel sizes is required to faithfully reach the Taylor-Aris dispersion regime. With the craft cutter used in this work we are able to cut a straight channel with length up to 20 cm (typically limited to a few centimeters in soft lithography due to the wafer size) and thus avoid the need for a serpentine channel \cite{bontoux2006,yan2015chip}.  


 Multi-layer soft lithography, which was first introduced by S.R. Quake and co-workers, is a technique to build an active microfluidic system containing microvalves and pumps \cite{11unger2000}. The system made from this technique consists of a flow channel and an air channel stacked perpendicular to each other, and the two channels are separated by a thin PDMS membrane. When pressurized air is applied in the air channel, the membrane is deflected and obstructs the flow in the flow channel resulting in a flexible actuator or valve without the need for manufacturing or control processes with resolution smaller than the scale of the channel itself.

Inspired by this multi-layer soft lithography technique, a novel design of double-layer parallel channels is presented within this work. Instead of placing the air channel and the flow channel perpendicular to each other, the channels are designed to be aligned parallel and exactly over top of one another. The parallel placement of two channels allows for direct control of the channel's cross-sectional shape by deforming one of the channel’s walls as depicted in Figure \ref{fig:4_ChipDesign}. The flow channel and the air channel are separated by a flexible membrane so that the membrane can bulge upward and downward in a controlled manner in response to the adjustable static pressure in the air channel.


\subsection{Deformable microchip fabrication}

\begin{figure*}
\centering
    \includegraphics[width=17.4cm]{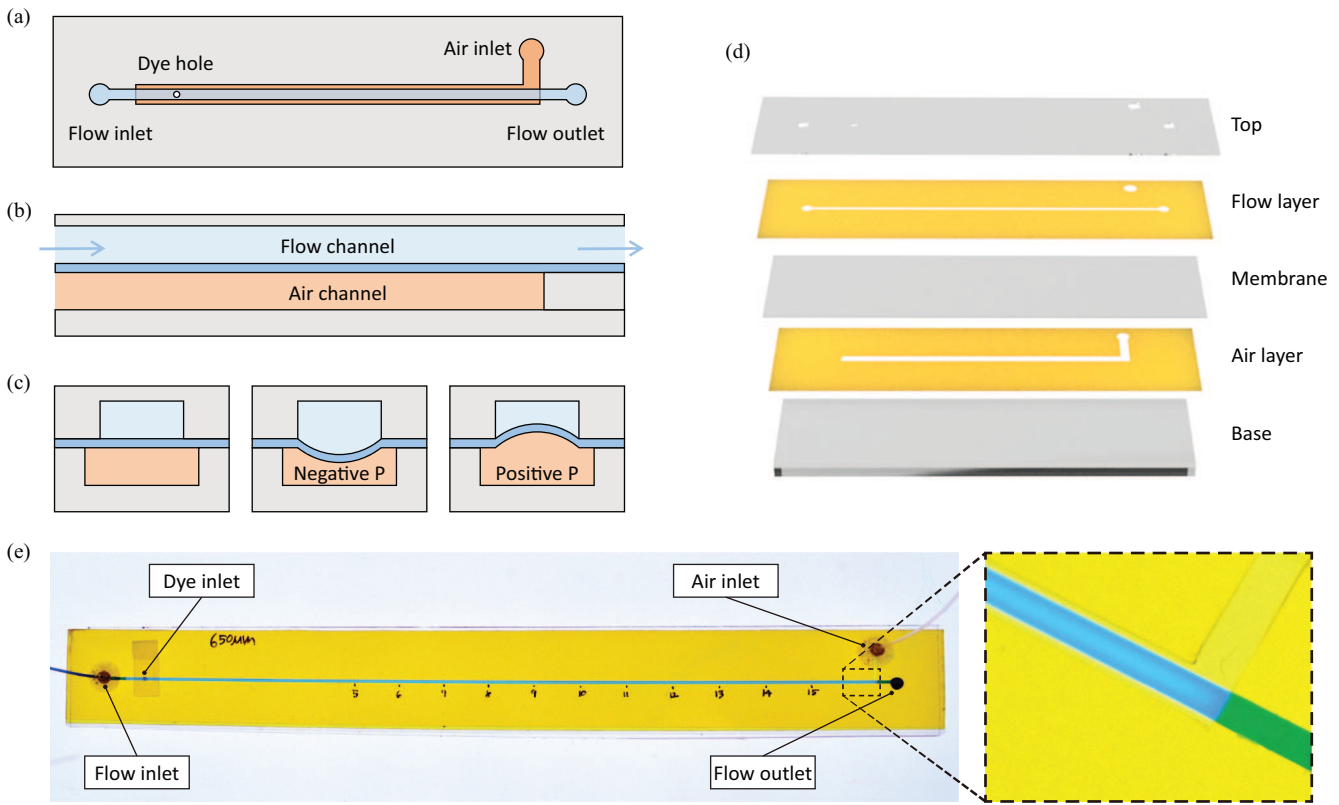}
    \caption{Double-layer deformable microchip design. (a - c) Schematic and sectional views of the microchip. The flow channel includes an inlet and an outlet while the air channel is a closed channel without an outlet. A flexible membrane placed between two channels is deformed via applied pressure (or vacuum) in the air channel. (d)  The complete chip is composed of five layers. Double-sided polyimide tape is used for the flow and the air layer. The top and the base layer are made of transparent PET film and acrylic, respectively. (e) Fabricated microchip. The deformed channel shape can be observed in the zoomed image; in this case, a positive pressure is applied to the air channel.  Blue food dye was used here for ease of visualization. }
    \label{fig:4_ChipDesign}
\end{figure*}

Each deformable microchip incorporates two channels which are placed parallel and directly over top of each other, with a flexible membrane between the two channels. Fluids flow along the straight flow channel and the air channel allows for control of the cross-sectional geometry of the flow by changing the pressure in the channel as shown in Figure \ref{fig:4_ChipDesign}.

For the flow layer and the air layer, double-sided polyimide tape (Bertech PPTDE-1) is used as the base material. This tape is of total thickness \SI{101.6}{\um} and is composed of a \SI{25.4}{\um} polyimide film with \SI{38.1}{\um} of silicone adhesive coating on each side. In the present work, we study three chips with different aspect ratios $\alpha = 3.0$, $4.8$, and $7.4$ corresponding to channel widths $w$ of \SI{300.4}{\um}, \SI{488.6}{\um}, and \SI{752.9}{\um}, respectively. 

To form an upper boundary of the flow channel, 0.1 mm thick trasparent PET film (Grafix Dura-Lar Clear Acetate Alternative) is cut into the chip size with four holes: two holes for the flow channel, one for the air channel, and one for introducing the tracer dye. A laser cutter (Universal Laser Systems 4.60 with 30 Watt laser) is used to cut the PET film. The small hole on the top layer is sealed with tape, and serves as direct access to the flow channel to introduce the tracer dye. We observed that the experimental results are insensitive to minor variations to the details of the initial dye patch we achieved using this direct dye introduction method, as documented in prior work for rectangular geometries \cite{8taylor2019}.

To construct the channels’ side walls, double-sided polyimide tape is cut by a craft cutter (Silhouette Cameo 3) to the desired channel design using the optimized protocol described in detail by Taylor \& Harris \cite{8taylor2019}. Once the cut is complete, the channel negative and the circular part for the air inlet is removed. The cut flow layer is placed over the top transparency and aligned with the inlet and the outlet holes on the transparency. The flexible membrane is attached on the tape, and the cut air layer is sealed on the membrane ensuring that two channels share the same centerline.  The minimum misalignment error that could be regularly achieved in this process was on the order of $15$ $\mu$m, although this step could take several attempts and became easier with practice.  A 3 mm thick optically clear acrylic sheet is attached on the air layer as a chip base ensuring mechanical stability and flatness. Capillary tubing (Cole Parmer PTFE \texttt{\#}30 AWG Thin wall tubing) and a tape gasket are adhered to the flow and air inlets using an epoxy adhesive (PARTS Express 5 Minute Quik-cure Epoxy adhesive).

\begin{figure*}[t]
    \centering
    \includegraphics[width=174mm]{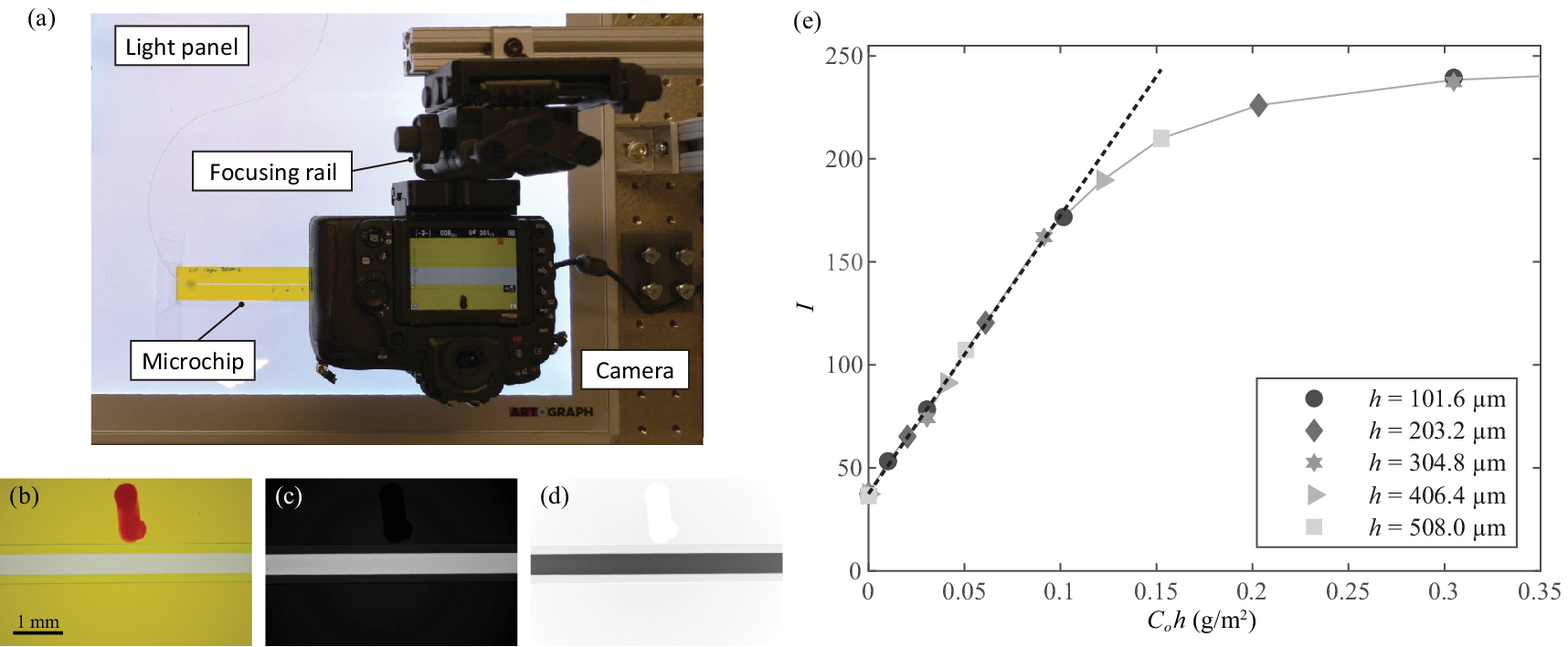}
    \caption{Setup and image processing for bright-field imaging. (a) Optical setup. (b) Raw image, (c) decomposed B channel, (d) inverted result from (c).  (e) Intensity linearity validation. Channels of different thickness $h$ (fabricated by changing the number of stacked double-sided tape layers) are filled with dye solutions of known concentration $C_o$ and the intensity $I$ of the inverted B channel is measured. The dashed line corresponds to a linear fit of $I = 137.3 C_oh+ 37.4$ up to an intensity of 160.  The value $I_o=37.4$ corresponds to the value measured when no dye is present ($C_o = 0$).}
    \label{fig:5_OpticalSetup}
\end{figure*}

\subsection{Membrane deformation}

A long and narrow rectangular membrane deformed under a uniform pressure loading takes an approximately 2D geometry for the majority of its length {\color{black} \cite{ziebart1998mechanical}}.  We will henceforth refer to this region of the membrane with a uniform cross-sectional shape as the ‘flat region.’  By avoiding the ends where 3D effects are observed (measured to be approximately 1 mm in length in the present work), the long flat region allows us to realize the experiment in a channel with a uniform but readily tunable cross-section. 
In this work, the entire experiment is conducted within the flat region of 15 cm length, always at least 5 mm away from the ends of the membrane.

In addition to the uniform pressure from the air channel, the fluid movement in the flow channel generates a pressure gradient along the channel length due to its viscosity. In this experiment, however, the pressure applied from the flow channel to the membrane can be safely neglected as the typical pressure applied from the air channel required to deform the channel significantly is on the order of 10 kPa whereas the pressure drop along the flow channel is in the order of 0.1 kPa (for the mean flow speeds of \SI{1.0}{mm/s} used in this work).  




For a given membrane material, a larger air pressure is needed to deform a narrower channel to the same deflection parameter.  Since a high pressure difference across the membrane was observed to result in air passing through the membrane, pressure differences higher than 200 kPa were not used in our experiments.  Considering the elastic behavior of the material and the applicable pressure range, plasticized PVC film (Stretch-tite premium plastic food wrap) is used for $\alpha=4.8$ and $7.4$ chips, and cross-linked silicon rubber film (Wacker ELASTOSIL Film 2030 250/50) is used for the narrowest chip at $\alpha=3.0$.

To apply pressure to the air channel, an empty syringe (filled with air at atmospheric pressure) is used and  connected to a check valve.  A pressure gauge is placed after the check valve and connected to the air channel by a capillary tubing. Once the channel reaches a desired pressure by filling or removing air in the system, the pressure is monitored to ensure there is no pressure change during the experiment.  {\color{black} In our experiments, the compliance (defined here as the maximum deflection per applied pressure) of the $\alpha=4.8$ and $7.4$ chips (food wrap) was approximately $\sim 1$ $\mu$m/kPa and the $\alpha=3.0$ chip (ELASTOSIL Film) was approximately $\sim 7$ $\mu$m/kPa.}

\subsection{Bright field imaging}

Fluorescence microscopy has been widely used in numerous fields of modern science including microfluidics due to its superior sensitivity and the linear relationship between the concentration of a fluorophore and its fluorescence intensity \cite{2,1}. However, the use of this technique requires a special excitation source such as a mercury lamp and a series of optical filters.  While the system has many advantages for measurements requiring high accuracy, a variety of issues have to be considered \cite{2}. One significant issue for the present experiments is the photobleaching effect which refers to a permanent loss of the ability for a fluorescent chemical compound to fluoresce. This effect can add significant error in the calibration between the intensity signal and the concentration field. Considering the apparent disadvantages of using fluorescence microscopy as well as overall higher equipment costs, bright field imaging on a flat light panel was used in the present work.


 The imaging setup is shown in Figure \ref{fig:5_OpticalSetup}(a). The microfludic chip is placed on 17 x 24 inch light panel (Artograph LightPad A950) to illuminate the microchip with uniform back lighting. To capture the intensity changes in the channel, a camera (Nikon D500) is placed vertically over the chip mounted on a focusing rail (Oben MFR4-5 Macro Focusing Rail) for precise camera alignment and focusing. To assess contrast between the channel region and the yellow tape representing the channel walls, the RGB channels of channel images with and without flurescein dye were decomposed and compared.  Among the three channels, the Blue (`B') channel demonstrated the best ability to distinguish the channel region.  In fluorescence microscopy, higher intensity represents a higher concentration of dye molecules along the light path. In the contrary, in bright field imaging, a larger number of dye molecules block the light arriving to the camera sensors which yields lower intensity. Hence, the raw B channel intensity values are first inverted by subtracting them from the maximum possible intensity value 255 to arrive a final map of the intensity $I$.  This sequence of image processing steps is shown for a sample channel in Figure \ref{fig:5_OpticalSetup}(b-d).

Images processed as above were obtained to verify and characterize a linear relationship between the measured intensity and the product of the channel thickness and a prescribed concentration of uniform dye solution in the channel.  This product (height times concentration) is proportional to the number of dye molecules within the cross-sectional area of the channel.  As shown in Figure \ref{fig:5_OpticalSetup}(e), a robust linear regime was found up to an intensity value of 160, after which the relationship becomes nonlinear.  By filling a channel with a known concentration of fluorescein dye solution, we can now use this relationship to measure the height profile of a deformed channel.  Furthermore, in this linear regime, we can also faithfully translate the intensity signal read by the camera into a 2D map of the dye concentration as it moves through the chip.  

\subsection{Dispersion experiment}
\begin{figure}
    \centering
    \includegraphics[width=84mm]{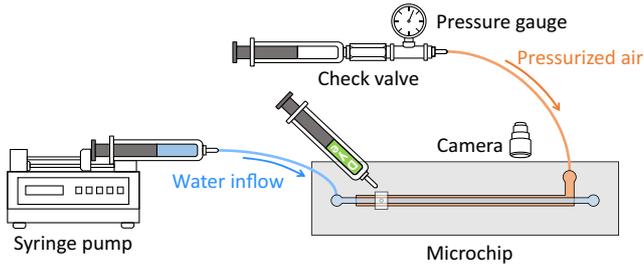}
    \caption{Schematic of the overall experimental setup. The pressure in the air channel is controlled using an air filled syringe, and monitored with a pressure gauge.  Fluid flow in the flow channel is introduced via a syringe pump.}
    \label{fig:6_ExpSetup}
\end{figure}

\begin{figure}[b]
    \centering
    \includegraphics[width=84mm]{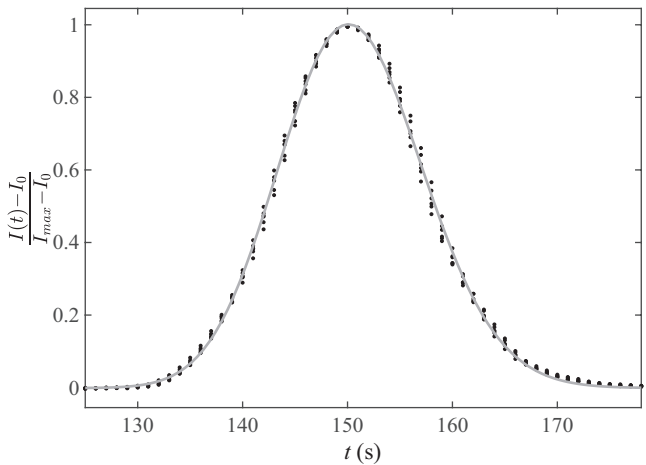}
    \caption{Curve fitting result for a sample experiment: $\alpha=7.4$ chip with $\delta=-0.36$.  The solid line is the best fit to equation (\ref{eqn:fit}) whereas the data points represent the results from six independent experimental trials.  The Taylor-Aris dispersivity $K$ is determined by the fit and can then be used to compute the dispersion factor $f$ using equation (\ref{eqn:f}).}
    \label{fig:7_CurveFii}
\end{figure}

The overall experimental setup is shown schematically in Figure \ref{fig:6_ExpSetup}. The fabricated microchip is connected to a glass syringe installed on a syringe pump (Harvard apparatus Standard Infuse-Withdraw Pump 11 Elite).  A macro lens (Mitakon Zhongyi 20 mm f/2 4.5x Super Macro Lens) is mounted on the camera to image the channel. The resolution of the images taken in this setup is 1.04 $\mu$m per pixel. The camera is centered 15 cm downstream from the dye inlet to ensure that the solute patch reaches the long-time asymptotic regime.  Prior work \cite{7dutta2001} has shown that in rectangular channel, the dispersivity approaches 95$\%$ of its long-time asymptotic value after a time 
\begin{equation}
t\approx\frac{1}{20}\frac{W^2}{\kappa}.
\label{eq:astimelimit}
\end{equation}
For the mean flow velocity of \SI{1.0}{mm/s} used throughout the present work, the system is estimated to be in the asymptotic regime after around 5 cm downstream, using equation (\ref{eq:astimelimit}) as a guide.

First, the deformed profiles of the flow channel are measured with various pressures applied to the air channel. By filling up the flow channel with dye solution of a known concentration, the deformation profile is measured using the linear relationship between the intensity and the height of the channel. The cross-sectional area of the channel is then calculated by numerically integrating the deformed channel profile. After this measurement, the channel is flushed with pure water to remove any residual dye solution. 

Once the flow syringe and flow channel are filled with water (and no air bubbles), the desired pressure is applied to the air channel. Any water exiting the outlet of the flow channel is wicked away using a lint-free cloth.

Next a small droplet of fluorescein (HiMedia Laboratories Fluorescein sodium salt) solution with a known molecular diffusivity $\kappa=$ \SI{5.7e-6}{\cm^2/\s} \cite{17aminian2016} is placed on the dye hole using a syringe full of fluorescein dye solution. A solution of \SI{2}{g/L} fluorescein dye is used which was confirmed to remain in the linear measurement regime at the downstream measurement location while also producing enough intensity to maximize the signal to noise ratio in the intensity signal. The dye hole is tape sealed again ensuring no bubble is formed near the hole. The flow rate required to generate a flow velocity of $U=$ \SI{1.0}{mm/s} is set on the syringe pump which is calculated based on the measured cross-sectional area, as described previously.

\begin{figure*}[b]
    \centering
    \includegraphics[width=174mm]{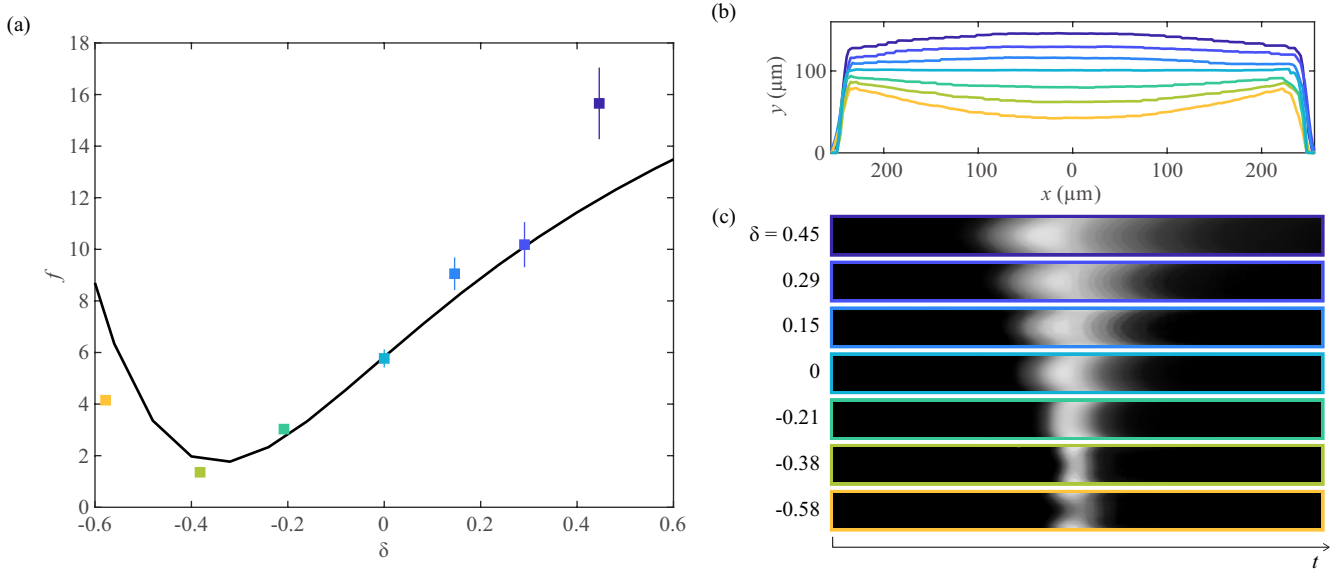}
    \caption{(a) Dispersion factor $f$ versus deflection parameter $\delta$ in the $\alpha=4.8$ chip. Numerical predictions of dispersion factors are shown as a solid black line, and experimental data are depicted with error bars drawn based on a standard multivariable error analysis \cite{8taylor2019}. (b) Measured cross-section of deformed {\color{black} flow} channel in experiment. (c) Intensity profile recorded at the fixed downstream position over time.  The intensity is normalized so that the maximum is the same brightness in all images.  In reality, the wider distributions (corresponding to high dispersion) are significantly more diluted.}
    \label{fig:8_4.8Result}
\end{figure*}

With a mean flow velocity of \SI{1.0}{\mm/\s} in these chips, the corresponding Peclet number $Pe$ is calculated to be $\sim 200$.  Photos are taken with an interval of 1 s at the fixed position 15 cm downstream from the dye hole.  In post-processing, for each photo, an average intensity value is computed over a cropped window spanning the entire width of the channel and 100 pixels ($\sim0.1$ mm) along its length.  This allows for the determination of a representative intensity value $I(t)$ as the dye plug passes through the imaging location.

For each geometry, intensity profiles $I(t)$ gathered from six independent trials are overlaid, and the collected intensity time series are curve fit to a translating and diffusing Gaussian distribution
\begin{equation}
I(t)-I_o=\frac{B}{\sqrt{4\pi Kt}}\exp\left[\frac{-(x_0-Ut)^2}{4Kt}\right]
\label{eqn:fit}
\end{equation}
as described in prior work \cite{8taylor2019}. The measurement position $x_0$ and the average velocity $U$ are known experimental parameters and so only the dispersivity $K$ and overall amplitude $B$ is fit during this process.  The overall amplitude $B$ varies between trials depending on the exact amount of dye introduced in the initial condition, but is otherwise irrelevant.  An example of such a fit is shown in Figure \ref{fig:7_CurveFii}.  In this plot, each data set is normalized by the peak measured intensity value $I_{max}$ so the curves can be superposed directly. Now knowing $K$ for our trial, the dispersion factor $f$ can thus ultimately be calculated by inverting equation (\ref{eqn:dispersivity_arbitrary}) yielding
\begin{equation}
f=\frac{210}{Pe^2}\left(\frac{K}{\kappa} -1\right).
\label{eqn:f}
\end{equation}

\section{Results}
Dispersivity of fluorescein dye in a deformed microchips with three different aspect ratios was tested. The measured widths of the flow channels are 300.4 \si{\um}, 488.6 \si{\um} and 752.9 \si{\um}, and the corresponding aspect ratios of the undeformed ($\delta=0$) chips are $\alpha=$3.0, 4.8 and 7.4, respectively.

Figure \ref{fig:8_4.8Result}(a) shows the measured dispersion factor $f$ with different deflection parameters $\delta$ in the $\alpha=4.8$ chip. The experimental results are compared with corresponding numerical predictions, described in Section 2.  Note that the theoretical curve assumes an idealized geometry with a fixed side wall height whereas in experiment the sidewalls also deform with the applied pressure, as can be seen in the measured channel geometries shown in Figure \ref{fig:8_4.8Result}(b).  Despite this difference, the overall agreement is very good.  In particular, our results demonstrate that precise control of the dispersion is achievable in a single microchannel, with the dispersion factor varying by nearly an order of magnitude between geometries.  Furthermore, the postulated presence of a local minimum in the dispersion coefficient has also been confirmed, and occurs at an inward bowing of around $\delta\approx -0.4$ for this microchip.


Since we used a fixed mean flow velocity of \SI{1.0}{mm/s} in all experiments, the trend in the dispersion factor is directly correlated to the length of the solute band in the flow direction. Figure \ref{fig:8_4.8Result}(c) shows the intensity profile over time at our measurement cross-section, corresponding to the data from Figure \ref{fig:8_4.8Result}(a).  Note that the figure is a time series of the intensity profile measured at a fixed spatial location, rather than an image of the solute distribution at a fixed moment in time.

\begin{figure}
    \centering
    \includegraphics[width=84mm]{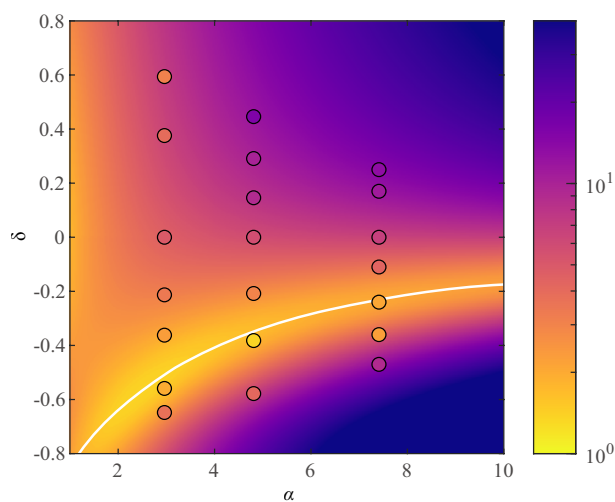}
    \caption{Experimental results of the dispersion factor $f$ for three different microchips corresponding to $\alpha=3.0, 4.8$ and $7.4$.  Results are overlaid on the numerical predictions. Each data point is derived from six independent experimental trials.  In this figure, $f$ values obtained in the simulation larger than 40 are depicted in the darkest color of the color map for improved contrast within the region of interest.}
    \label{fig:9_FullExpResult}
\end{figure}

Figure \ref{fig:9_FullExpResult} summarizes the results obtained from all three chips fabricated in the present study.  In general the comparison to the idealized theory is good, and each shows a local minimum in the dispersion factor for a particular level of inward bowing, as predicted numerically and in prior numerical and theoretical work by Dutta et al. \cite{3dutta2006}.  Some discrepancies and limitations to the comparison do exist, and will be detailed in the following section.


\section{Discussion} 

While the overall trends of the numerics are well captured by our experimental results, some quantitative discrepancies do exist.  In this section we will discuss some of the limitations of the comparison and approach.

The numerical values are calculated based on an idealized deformed channel shape that is defined by only two parameters: the aspect ratio and the deflection parameter, where the upper arc of the bowed wall is assumed to be strictly parabolic.  However, the channel geometry measured from the experiment shows that this assumed geometry in the numerical calculation is of course not a perfect representation.  This effect leads to the largest discrepancy in our comparison for the smallest aspect ratio chip when $\delta>0$, as can be seen in the upper left corner of Figure \ref{fig:9_FullExpResult}.  A better quantitative agreement can be obtained by inputting the measured channel shapes directly into simulation, although this does come at the cost of introducing a much larger set of geometric parameters.

It is also possible that the comparison could be improved if the side wall height of the channel remained fixed, independent of $\delta$.  In our experimental work we measured a clear deformation of the side walls that was correlated with the applied pressure in the air channel.  Different material selection for the flow channel, mechanical clamping of the overall chip, {\color{black} or a different manufacturing method} could potentially mitigate this effect.

Lastly, in long-time regime, the intensity profile of the dye patch along the channel direction should converge towards a simple Gaussian distribution. However, we observed notably skewed intensity profiles for cases with positive deflection parameters $\delta>0$. The profile becomes increasingly skewed with larger deflection parameter $\delta$ and is likely another factor contributing to the quantitative discrepancy between the numerics (computed from long-time asymptotic equations) and experiments.  
In designing the experiment, the value computed from equation (\ref{eq:astimelimit}) was used as a criterion to determine the minimum distance the solute patch should travel before measuring the intensity profile.  However, this guideline was only verified numerically in rectangular channels, which is evidently different for the deformed channel shape considered in our work.  Future theoretical and numerical work might consider other methods to better quantify the evolution of transient asymmetries in the distribution captured by higher moments (such as the skewness), as was recently documented for rectangular and elliptical geometries \cite{17aminian2016}.


\section{Conclusions}
We have established an accessible method to control the dispersion factor within a single multi-layer microchannel by deforming one of the channel walls in a controlled manner.  The fabrication technique uses a commercial craft cutter which allows us to achieve this goal without relying on the expensive equipment and facilities required by other common microfabrication techniques.  The dispersion control technique presented here does not rely on manufacturing or control process with resolution beyond that of the microchannel itself.  To complete dispersion measurements in our custom channels, we developed and validated a robust bright-field imaging setup using an artist's tracing light pad and DSLR camera, which further reduces the cost for performing such experiments.

In our experimental measurements, we were able to isolate the sensitive relationship between the dispersion factor $f$ and the deflection parameter $\delta$ on a single microchip.  In particular, for three microchannels of different aspect ratios, we experimentally verified the presence of a local minimum in the dispersion factor as $\delta$ is varied continuously.  This minimum value of $f$ was found when the upper channel wall was deformed inward by an amount which depends on the overall aspect ratio of the channel.

Future work may include more detailed quantitative assessment of the early time distributions in the experiment where significant skewness in the distribution is present \cite{17aminian2016}. Furthermore, the numerical code could be paired with an optimization scheme to identify other channel geometries with even lower dispersion coefficients.

Given that the channel cross-section (and thus the dispersion factor) can now be controlled on a single chip, our contributions open up potential for other exciting areas of study.  One area of interest might be in implementing real-time {\it active} control of dispersion processes in pressure-drive{\color{black}n} flows. Alternatively, a similar device to ours could be used to study dispersion through a time-dependent channel geometry \cite{marbach2019} with potential relevance to transport in biological processes.

\section*{Conflicts of interest}
There are no conflicts to declare.

\section*{Data availability}
The datasets generated during and/or analysed during the current study are available from the corresponding author on reasonable request.

\begin{acknowledgements}
DMH and GL gratefully acknowledge the financial support of the Brown OVPR Salomon Award. JLM and AL were supported in part by NSF CAREER Grant DMS-1352353 (2014-2020) and NSF Applied Math Grant DMS-1909035 (2019-Present). Furthermore, GL and DMH would like to acknowledge A. Taylor for support and guidance with the craft-cutter technique, K. Dalnoki-Veress for advice on membrane selection, and K. Breuer for use of his inverted microscope in preliminary experiments.
\end{acknowledgements}



\balance

\bibliographystyle{spmpsci}      
\bibliography{Mybib} 

\end{document}